\documentclass[aps,pre,twocolumn,groupedaddress,showpacs]{revtex4}
\usepackage{graphicx}
\begin{document}

\title{Interplay between phase ordering and roughening on growing films}

\author{Barbara Drossel}
\affiliation{Institut f\"ur Festk\"orperphysik, TU Darmstadt,
Hochschulstr.~6, 64289 Darmstadt, Germany }

\author{Mehran Kardar}
\affiliation{Department of Physics, Massachusetts
Institute of Technology, Cambridge, Massachusetts 02139}

\date{\today}

\begin{abstract}
We study the interplay between surface roughening 
and phase separation during the growth of binary films.
Renormalization group calculations are performed on a pair of equations coupling the
interface height and order parameter fluctuations.
We find a larger roughness exponent at the critical point of the order parameter compared to the disordered phase,
and an increase in the upper critical dimension for the surface roughening transition
from two to four.
Numerical simulations performed on a solid-on-solid model with two types of
deposited particles corroborate some of these findings.
However, for a range of parameters not accessible to perturbative analysis,
we find non-universal behavior with a continuously varying dynamic exponent.
\end{abstract}


\maketitle
  
\section{Introduction}
\label{intro}

Thin solid films are grown for a variety of technological
applications, using molecular beam epitaxy (MBE) or vapor deposition.
In order to create materials with specific electronic, optical, or
mechanical properties, often more than one type of particle is
deposited.  When the particle mobility in the bulk is small, surface
configurations become frozen in the bulk, leading to anisotropic
structures that reflect the growth history, and are different from
bulk equilibrium phases. If, for instance, a combination of particles
are deposited that tend to phase separate at the surface, the grown
films have lamellae or columns of the two phases that extend parallel
to the growth direction \cite{dev92,ada92}. This process of phase
separation, as well as other ordering phenomena, can be affected by
elastic forces, by the orientation of the growing crystal, by
properties of the substrate, and by surface roughness. The range of
possible scenarios is very rich and far from understood. There
are a variety of analytical and computer models which try to shed
light on some of these phenomena, but a systematic study and
understanding of the possible phase transitions does not yet exist.

In this paper, we focus on the interplay between phase separation
and surface roughening, neglecting the possible influence of elastic
forces, substrate properties, and orientation dependencies due to the
crystal structure. There exist several theoretical studies of phase
separation during growth that neglect also the effect of surface
roughness.  In all these models it is assumed that the mobility of the
atoms in the bulk is zero, such that all of the dynamics occurs at the
surface.  A model in which the probability that an incoming atom
sticks to a given surface site depends on the state of the neighboring
sites in the layer below \cite{kan90}, leads to a phase separation
transition in the universality class of ordinary Ising models, if the
model is symmetric with respect to the two phases. The same
conclusion applies a model in which the top layer is
fully thermally equilibrated before the next layer is added
\cite{dro97}. A model for spinodal decomposition during growth was
introduced in Ref.~\cite{atz92}. 
In this model, phase separation is due to
surface diffusion, and is limited due to the current of incoming
particles, leading to a characteristic scale for the thickness of
lamellae or columns, as confirmed by Monte-Carlo
simulations \cite{ada93}.

However, the layer by layer growth mode underlying these models is
unstable, and growing surfaces generally are rough. Several studies
exist that investigate growth models that contain both phase
separation and surface roughness. Simulations of an Eden model with
two types of particles suggest that the surface roughness increases
due to the phase separation \cite{aus93}. A solid-on-solid growth
model where the adsorption probabilities for the two types of
particles depend on the local neighborhood in the layer below leads
also to an increased surface roughness \cite{kot98}. The reason is
that particles are more likely to be adsorbed within domains than at
domain boundaries. On length scales much larger than the domain size,
a crossover to the scaling behavior of the Kardar-Parisi-Zhang (KPZ)
equation \cite{kar86} is found. Another computer model
where particles are adsorbed randomly and subsequently diffuse along
the surface leads to domains whose thickness is a nonmonotonous
function of the deposition rate and the temperature, and for a certain
range of parameter values, the height profile has steep steps at
domain boundaries \cite{leo97a}. A set of coupled Langvin equations
for this model is suggested in Ref.~\cite{leo97b} and studied using
stability analysis and Fourier decomposition. 

These studies are rather incomplete, and in particular lacking a
discussion of the possible effects of the height profile on the phase
separation dynamics.  A first attempt to a systematic study of the
possible phases and scaling behaviors of coupled phase separation and
roughening during growth was presented in a recent letter by us
\cite{ourprl}. A set of two coupled Langevin equations was suggested,
and computer simulations in 1+1 dimensions were performed, revealing a
rich phase diagram. It is the purpose of this paper to extend and
deepen this short study by presenting a renormalization group (RG)
analysis, and further simulation results. While the RG analysis gives
information about the behavior of the system in dimensions close to
4+1 and higher, computer simulations are particularly efficient in low
dimensions. The general results obtained by the two approaches are
compatible with each other. In addition, the computer simulations in
1+1 dimension reveal interesting nonuniversal behavior for a range of
parameters that cannot be studied using perturbative RG.  

The outline of the paper is as follows: In section \ref{equations}, we
introduce and discuss the coupled set of Langevin equations used in
this paper. Scaling laws and critical exponents will also be
defined. Section \ref{RG} presents results of the RG analysis of these
equations. One of the main findings is that the lower critical
dimension for the surface roughening transition is increased from 2 to
4 dimensions due to the coupling to the critical phase ordering
dynamics. Section \ref{simulations} presents results of computer
simulations.  Section \ref{other} analyses the connection of our model
with the advection of a passive scalar in a velocity field, and with
directed polymers drifting through a random medium. Section
\ref{conclusions} contains a summary and discussion of our results.

\section{Equations of motion and scaling laws}
\label{equations}

We consider the growth of a binary alloy on a $d$-dimensional
substrate.  Let ${\bf x}$ be the coordinate perpendicular to the
growth direction, and $t$ the time. Since we assume that all dynamics
occurs at the surface of the growing material, the equations of motion
can be expressed in terms of ${\bf x}$ and $t$ alone.  In order to
characterize surface roughness and phase ordering, we introduce the
height variable $h({\bf x},t)$, which is 
the surface profile at position ${\bf x}$ at time $t$, 
and an order parameter $m({\bf x},t)$, which is the difference
in the densities of the two particle types at the surface at position
${\bf x}$ and time $t$.  The interplay between the fluctuations in
$m$, and the height $h$ is captured phenomenologically by the coupled
Langevin equations,
\begin{eqnarray}
\partial_t h &=& \nu \nabla^2 h + {\lambda\over 2} (\nabla h)^2 +
{\alpha\over 2} m^2 +\zeta_h, \label{langevin1}\\ \partial_t m &=& K
(\nabla^2m-rm-um^3) + a\nabla h \cdot \nabla m+bm\nabla^2 h\nonumber \\
&& + {c\over 2} m (\nabla h)^2 + \zeta_m ,\label{langevin2}
\end{eqnarray}
with
\begin{eqnarray*}
\langle \zeta_h({\bf x},t) \zeta_h({\bf x}',t')\rangle&=&2D_h \delta^d({\bf x} 
- {\bf x}')\delta(t-t'),\\
\langle \zeta_m({\bf x},t) \zeta_m({\bf x}',t')\rangle&=&2D_m \delta^d({\bf x} 
- {\bf x}')\delta(t-t').
\end{eqnarray*}

Since we are interested in the critical behavior of the model, we have
assumed that it has the symmetry $m \to -m$, and  included all
potentially relevant terms compatible with this symmetry. In
experiments or computer simulations, this symmetry can be achieved by
tuning the ratio between the two types of adsorbed particles to the
appropriate value. In the absence of such an order parameter symmetry, the
system  may undergo a first-order phase transition which is not considered
here. Equation (\ref{langevin1}) is the Kardar-Parisi-Zhang (KPZ)
equation \cite{kar86} for surface growth, plus the leading coupling to
the order parameter.  Equation (\ref{langevin2}) is the time dependent
Landau--Ginzburg equation for a (non-conserved) Ising model, with
three different couplings to the height fluctuations.  The Gaussian,
delta-correlated noise terms, $\zeta_h$ and $\zeta_m$, mimic the
effects of faster degrees of freedom.

These equations apply to growth by vapor deposition, with particles
sticking at surface sites with a probability that depends on the local
environment in the growing film. The coupling terms in the Langevin
equations (\ref{langevin1}) and (\ref{langevin2}) have obvious meaning
in this context: The term proportional to $\alpha$ implies that particles are more
likely to be absorbed within domains where they feel a stronger
attractive force (if $\alpha >0$). A negative $\alpha$ can also be
meaningful: if the adsorption rate within domains is limited by the
availability of particles of the correct type, this can slow down
growth. However, if this is due to the vapor phase not being well
stirred, additional equations for the particle concentrations in the
vapor phase will be needed. Such equations are not included in this
paper. The contribution from $a$ (with $a>0$) implies that domain walls tend to
be driven downhill; e.g. if the identity of a newly adsorbed
particle is more likely to be affected by its uphill neighbors than by
the downhill ones. A positive $b$ indicates that new domains are
more likely to be formed in hilltops where there are less neighbors
that could influence the type of particle to be adsorbed. The term proportional to
$c$ is similar in character to the KPZ nonlinearity $\lambda$, and means
that susceptibility to phase separation depends on the slope.

Models for MBE typically assume
that particle deposition at the surface is random, and that no
desorption of particles takes place. In this case, the height profile
and the order parameter dynamics are shaped by diffusion of particles
along the surface. This physical situation leads to a different set of
Langevin equations,
\begin{eqnarray}
\partial_t h &=& \nu \nabla^2 h + {\beta\over 2}  \nabla^2 m^2 + \zeta_h, 
\label{langevin3}\\ 
\partial_t m &=& K\nabla^2m - v m 
+ \zeta_m ,\label{langevin4}
\end{eqnarray}
where we have again imposed the symmetry $m \to -m$.
The noise terms $\zeta_m$ and $\zeta_h$ have the same
nonconserved correlations as for the vapor-deposition model above, due
to the incoming particle current. Because of the conservation of
volume during surface diffusion, the
deterministic terms on the right-hand side of Eq.~(\ref{langevin3})
must be the divergence of a current, disallowing the terms proportional to
$\lambda$ and $\alpha$ in Eq.~(\ref{langevin1}). A
negative value of $\beta$ means that particles are more likely to be
adsorbed within domains if they are not needed to the same extent in
the neighborhood.  The equation for the order parameter has also the
form of the divergence of a current (we have only included the
lowest-order term), plus a nonconserved contribution $-vm$ due to the
incoming current of particles that tends to reduce the value of the
order parameter.  The lowest order coupling to the height variable in
Eq.~(\ref{langevin4}) is of the form $\nabla^2(m\nabla^2 h)$,
which is irrelevant.  In contrast to the vapor deposition case, the
parameter $v$ can never become negative, making a term in $m^3$ 
unnecessary.  Instead, the tendency towards phase separation is
locally captured by $K<0$, which is the
main focus of the work by L\'eonard and Desai \cite{leo97a,leo97b},
and of Atzmon et al \cite{atz92} (the latter study does not include
the height profile). 
A change in the sign of $K$ marks the onset of phase separation
in models of conserved dynamics.
Higher order terms, such $\nabla^4m$ or $\nabla^2 m^3 $ which are 
included by other authors, are then needed for the stability of short
wavelength fluctuations, and may also affect the precise shape of 
the order parameter profile within domains.
Such terms are irrelevant to considerations of the long wavelength
behavior of Eq.~(\ref{langevin4}), since in this case instabilities
can only persists up to a length scale of the order $\sqrt{K/v}$, set by
the current of incoming particles.  
Since higher temperatures and deposition rates favor mixing, it
is likely that $K$ eventually becomes positive as these parameters are increased.
If the substrate dimension is $d=1$ (or is effectively $d=1$ because
diffusion proceeds along a preferred direction) the
coarse-grained value of $K$ must be positive,
as the dynamics are then similar to a 1-dimensional Ising model, which
cannot have an ordered phase.  For this reason, the choice of
$K<0$ in \cite{leo97a,leo97b} does not capture the long wavelength
behavior of the system.

The main focus of the next two sections is on the model for vapor
deposition, Eqs.~(\ref{langevin1}) and (\ref{langevin2}), and we
discuss Eqs.~(\ref{langevin3}) and (\ref{langevin4}) only briefly in
connection with the RG calculation. Our analysis of the models will
concentrate on the scaling behavior of the height profile and of the
order parameter. On sufficiently large length scales, height profiles
of growing interfaces are usually characterized by a scaling form
\begin{equation}\label{Chh}
\langle \left[h({\bf x},t)-h({\bf x'},t')\right]^2\rangle 
\sim |{\bf x-x'}|^{2\chi} g\left(|t-t'|\over |{\bf x-x'}|^{z_h}\right),
\end{equation}
where $\chi$ is the roughness exponent, and $z_h$ is a dynamical
scaling exponent. The values of the exponents depend on the underlying
growth model, and one of our objectives is to find out how they are
affected by the coupling to the order parameter dynamics. 

The scaling of the order parameter is different along the growth direction
and perpendicular to it. In contrast to the height variable, the order
parameter is unlikely to be exactly at a  fixed point, and for
this reason we include a correlation length $\xi$. We also have to
allow for the possibility that the height and the order parameter
dynamics have different dynamical critical exponents $z_h$ and $z_m$.
The scaling laws for the order parameter then read
\begin{eqnarray}
G_m^{(x)}({\bf x-x'})&\equiv& \langle m({\bf x},t) m({\bf x'},t) \rangle \nonumber\\
&=&|{\bf x-x'}|^{\eta-1}g_m^{\perp}(|{\bf x-x'}|/\xi)\nonumber\\ 
G_m^{(t)}(t-t') &\equiv& \langle m({\bf x},t)m({\bf x},t')\rangle\nonumber\\
&=& |t-t'|^{(\eta-1)/z_m} g_m^{\parallel}(|t-t'|/\xi^{z_m})\, . \label{corr}
\end{eqnarray}

\section{Renormalization group analysis}
\label{RG}

Let us now renormalize the equations of motion, Eqs.~(\ref{langevin1})
and (\ref{langevin2}), and search for fixed points that are accessible
by perturbation theory. Inserting the equations of motion in the
Gaussian probability distribution of the noise
\begin{equation}
W[\zeta_h,\zeta_m] \propto \exp\left\{-\int d^d x ~ dt \left[ \frac{\zeta_h({\bf x},t)^2}{4D_h}+\frac{\zeta_m({\bf x},t)^2}{4D_m}\right]\right\}\, , \label{noise}
\end{equation}
and introducing auxiliary fields $\tilde m$ and  $\tilde h$, we obtain the weight of a given space-time configuration $[h({\bf x},t),m({\bf x},t)]$ \cite{BJW}
\begin{displaymath}
W[h,m] \propto \int {\cal D} [i\tilde h] \int {\cal D} [i\tilde m] \exp\left\{{\cal J} [\tilde h,h,\tilde m,m]\right\}\,,
\end{displaymath}
with the dynamical functional
\begin{eqnarray}
{\cal J}&[&\tilde h, h,\tilde m,m] =\int d^d x \int dt \Biggl\{D_h \tilde h \tilde h \nonumber -\tilde h \times\\
&&\times \left[{\partial h\over \partial t} - \nu \nabla^2 h - {\lambda \over 2} (\nabla h)^2 - {\alpha \over 2} m^2  \right] \nonumber\\
&& +D_m \tilde m\tilde m -\tilde m\times\\
&&\times \Bigl[{\partial m\over \partial t}-K(\nabla^2 m -rm-um^3)\nonumber\\
&& -a\nabla h \cdot \nabla m - bm\nabla^2 h-\frac{c}{2}m(\nabla h)^2 \Bigr]
\Biggr\}\,.
\label{functional}
\end{eqnarray}
The dynamical functional ${\cal J}$ plays the same role in dynamical
RG as the Hamiltonian in statics. The bare propagators of this model are
\begin{equation}
G^h_0({\bf k}, t) \equiv \langle \tilde h(-{\bf k}, t) h({\bf k},t) \rangle_0 = \theta(t) e^{-\nu k^2 t}\, , \label{resph}
\end{equation}
\begin{equation}
C^h_0({\bf k}, t) \equiv \langle h(-{\bf k}, t) h({\bf k},t) \rangle_0 = \frac{D_h e^{-\nu k^2 |t|}}{\nu k^2} \, ,\label{corrh}
\end{equation}
\begin{equation}
G^m_0({\bf k}, t) \equiv \langle \tilde m(-{\bf k}, t) m({\bf k},t) \rangle_0 = \theta(t) e^{-K (r+k^2) t}\, , \label{respm}
\end{equation}
\begin{equation}
C^m_0({\bf k}, t) \equiv \langle m(-{\bf k}, t) m({\bf k},t) \rangle_0 = \frac{D_m e^{-K (r+k^2) |t|}}{K(r+k^2)} \, ,\label{corrm}
\end{equation}
and the interaction vertices are obtained from the higher-order terms in $J$.  In
the diagrams below, $h$ and $\tilde h$ are represented by straight and
wiggly lines respectively. Lines for the order parameter
$m$ are represented the same way, with an additional short dash
perpendicular to the propagator. 

The bare dimensions $d_0$ of the couplings are obtained by rescaling
space, time, height, and order parameter according to $x=bx',~t=b^z
t',~h=b^\chi h',~\tilde h=b^{\tilde \chi} \tilde h',~m=b^\zeta m',~
\tilde m=b^{\tilde \zeta} \tilde m'$, and by requiring invariance of
the Gaussian part of 
${\cal J}$ under such rescaling. (The scaling dimension $\zeta$ of the order
parameter is related to the exponent $\eta$ defined in
Eq.~(\ref{corr}) via $\zeta=(\eta-1)/2$.) The results are listed in
table \ref{tab1}. In the following, we analyze the scaling
behavior resulting from an RG analysis as function of the spatial
dimension $d$. 

\subsection{Dimensions $d>6$}

In sufficiently high dimensions, the Gaussian fixed
point, which is characterized by uncoupled, linear Langevin equations
is stable with respect to the higher-order terms. The condition of
scale invariance of the linear Langevin equations leads to $z=2$ and
$\chi=\zeta=(2-d)/2$, and to the scaling dimensions $d_g$ listed in
the third column of table \ref{tab1}. In dimensions $d>6$, all
nonlinear couplings are irrelevant. The surface is smooth, and the
order parameter goes through a classical phase transition. 

\subsection{Dimensions $4<d<6$}

\begin{figure}
\includegraphics*[width=2cm]{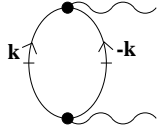}
\caption{The diagram renormalizing $D_h$. The lines with a small
bar represent the order parameter propagators, while lines
without this bar belong to the height variable. Wiggled lines stand
for $\tilde h$ and $\tilde m$, and smooth lines for $h$ and $m$. \label{fig1} 
}
\end{figure}

Below $d=6$, the coupling $\alpha$ becomes relevant. The Gaussian
fixed point still exists, but becomes unstable. A new stable fixed
point with a nonzero value of $\alpha$ emerges. Whenever nonlinear
terms cannot be neglected, the couplings change under rescaling not
only according to their bare dimensions, but also according to those
contributions that are generated under renormalization.
Renormalization of this model is done by first integrating over the
large wave vectors $\Lambda/b < k < \Lambda$, where $\Lambda$ is the
wave vector cutoff, and the scaling factor $b$ is larger than 1. Next,
the system is rescaled to the original size by introducing new
variables $k' = bk$, $t' = t/b^z$.  This procedure involves an
expansion of $e^{\cal J}$ in the couplings.  In this way, the coupling
$\alpha$ generates a contribution to $D_h$, which is graphically
represented by the diagram in Fig.~\ref{fig1}.
Evaluation of this diagram gives a contribution to $D_h$ of
\begin{eqnarray*}
B&=&\frac{\alpha^2D_m^2}{2K^2}\int\limits_{\Lambda/b < |{\bf k}| < \Lambda} d^d k \int\limits_0^\infty dt \frac{e^{-2K(r+k^2)|t|}}{(r+k^2)^2}\\
&=& \frac{\alpha^2D_m^2K_d\Lambda^{d-6}(1-b^{6-d})} {4K^3 (d-6)},
\end{eqnarray*}
where $K_d$ is the surface of the
$d$-dimensional unit sphere, divided by $(2\pi)^d$. 
We have also set $r=0$,
assuming that the order parameter is exactly at its critical point.
The renormalized value of this parameter is thus 
$$D_h'=b^{z-d-2\chi}[D_h+B] \, .$$ 
Setting $b=1+dl$, we obtain the flow equation
\begin{equation}
\frac{dD_h}{dl} = -D_h(d+2\chi-z)+\frac{\alpha^2 D_m^2K_d \Lambda^{d-6}}{4K^3}\, .\label{flowdh}
\end{equation}

The exponents $z$ and $\zeta$ are fixed at the values $z=2$ and $\zeta =(2-d)/2$,
since the  renormalization of the parameters $\nu$, $K$, $D_m$ does not obtain 
any anomalous contributions from diagrams. 
The condition that $\alpha$ has a nonzero fixed point leads to $\chi=4-d$. 
With these values of the exponents, the condition that $D_h$ is invariant under 
rescaling leads to the fixed point value of $D_h$
\begin{equation}
D_h = \frac{\alpha^2 D_m^2K_d \Lambda^{d-6}}{4K^3(6-d)}.
\label{Dalpha}
\end{equation}
This fixed point, where $\alpha$ is the only nonzero
coupling is stable between 4 and 6 dimensions. The scaling dimensions
of the other couplings are given in the right-hand column of table
\ref{tab1}.

Note that for $4<d<6$, the term $\alpha m^2$ can be regarded
as  a correlated noise acting on the surface
height. This correlated noise is more relevant than the white noise,
and incrases the value of the roughness exponent $\chi$ from
$(2-d)/2$ to $4-d$. As this value is still negative, the surface is
flat at this fixed point. 

\begin{table}
\begin{tabular}{r|c|c|c}
coupling & $d_0$ & $d_g$ & $d_\alpha$ \\
\hline
$D_m$ & $z-d-2\zeta$ & 0 & 0 \\
$K$ & $z-2$ & 0 & 0 \\
$r$ & 2 & 2 & 2 \\
$u$ & $2+2\zeta$ &$ 4-d$ &$ 4-d$ \\
$a$ & $z-2+\chi$ & $(2-d)/2 $& $4-d$ \\
$b$& $z-2+\chi$ & $(2-d)/2$ &$ 4-d $\\
$c$ & $z-2+2\chi$ & $2-d $&$ 2(4-d)$ \\
$D_h$ &$ z-d-2\chi$ & 0 & $d-6$ \\
$\nu$ & $z-2$ & $0$ &$ 0$ \\
$\lambda$ & $z-2+\chi$ & $(2-d)/2$ &$ 4-d$ \\
$\alpha$ & $2\zeta -\chi + z$ & $(6-d)/2$ & 0 \\
$\beta $ & $2\zeta-\chi+z-2$ & $(2-d)/2$  
\end{tabular}
\caption{Bare dimensions $d_0$, scaling dimensions $d_g$ at the Gaussian fixed point, and scaling dimensions $d_\alpha$ between 4 and 6 dimensions of all the couplings.}\label{tab1}
\end{table}

\subsection{Dimensions $d \simeq 4$}
Below $d=4$ dimensions, the flat phase becomes unstable, because the
roughness exponent becomes positive, and the coupling $\lambda$
obtains a positive scaling dimension. In $d=4+\epsilon$ dimensions
(with $|\epsilon|$ small), we can therefore expect a fixed point where
$\lambda$ (or a power of $\lambda$) is of the order of $\epsilon$. 
In order to find this fixed point, let us first assume that there is no
feedback from the height to the order parameter ($a=b=c=0$), and then
take into account all terms of the lowest order in $\lambda$. 

\begin{figure}
\includegraphics*[width=7cm]{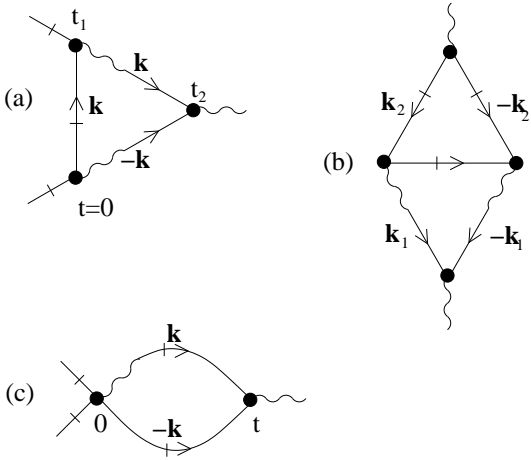}
\caption{The diagrams to be considered near $d=4$ dimensions.\label{fig2} 
}
\end{figure}

As shown in Figs.~\ref{fig2}(a) and \ref{fig2}(b), there
are two diagrams that contain one $\lambda$ vertex.
Diagram (a) makes a contribution to $\alpha/2$ equal to
\begin{eqnarray*}
A&=& \alpha^2\lambda \int\limits_{\Lambda/b < |{\bf k}| < \Lambda} d^d k \int\limits_0^\infty dt_1 \int\limits_{t_1}^\infty dt_2 \\ && \frac{k^2D_me^{-\nu k^2(2t_2-t_1)-K(r+k^2)t_1}} {K(r+k^2) }\\
&=& \frac{\alpha^2\lambda D_m K_4\Lambda^\epsilon  dl}{2K\nu (K+\nu)} \, , 
\end{eqnarray*}
which is a correction of 
order $\lambda\alpha^2$.  Fig.~\ref{fig2}(b) is a correction to
$D_h$ of order $\lambda\alpha^3$.  Since it modifies the flow equation
Eq.~(\ref{flowdh}) for  $D_h$
only to order $\epsilon$, it need not be
evaluated.  Furthermore, Eq.~(\ref{langevin2})
describes for $a=b=c=0$ the relaxational dynamics of an order
parameter in the universality class of the Ising model, which is known
to have a non-trivial stable fixed point for $D_mu/K=-K_4\epsilon/9 + {\cal
O}(\epsilon^2)$ below 4 dimensions, and with $r$ of the order of $\epsilon$
\cite{hh}. This means that we have to take into account additionally
diagram \ref{fig2}(c), which makes for $d<4$ a contribution 
$$C=-3\alpha u K \int\limits_{\Lambda/b < |{\bf k}| < \Lambda} d^d k
\int\limits_0^\infty dt \frac{D_me^{-2K(r+k^2)t}}{K(r+k^2)} =
\frac{\epsilon\alpha dl}{6}\, $$  
to $\alpha/2$, which is of order $\epsilon$. 
In evaluating this
expression, we have inserted the above-mentioned fixed point value of
$D_mu/K$ and have set $r=0$, considering only the leading
contribution in an expansion in $\epsilon$. 

Taking all these results together, we obtain the following set of flow equations to order $\epsilon$: 
\begin{eqnarray}
\frac{dD_m}{dl} &=& D_m(z-d-2\zeta)\, ;\nonumber\\
\frac{dK}{dl} &=& K(z-2)\, ;\nonumber\\
\frac{d\nu}{dl} &=& \nu(z-2)\, ;\nonumber\\
\frac{d\lambda}{dl} &=& \lambda(z-2+\chi)\, ;\nonumber\\
\frac{dD_h}{dl} &=& -D_h(d+2\chi-z)+\frac{\alpha^2 D_m^2K_d \Lambda^{d-6}}{4K^3} + {\cal O}(\epsilon)\, ;\nonumber\\
\frac{d\alpha}{dl} &=& \alpha\left[2\zeta-\chi+z+ \frac{\theta(-\epsilon)\epsilon}{3} + \frac{ \alpha\lambda D_m K_4}{K\nu(K+\nu)}\right].
\label{flowalpha}
\end{eqnarray}

From the flow equations for $K$, $\nu$ and $D_m$, we obtain again the fixed point condition $z=2$ and $\zeta=(2-d)/2$. 
For $\epsilon>0$, the fixed point $\lambda=0$ is stable, and we have a negative roughness exponent $\chi=4-d$, as before. 
For $\epsilon<0$, the fixed point $\lambda=0$ is unstable, with the roughness exponent $\chi$ modified due to diagram (c) in Fig.~\ref{fig2}. For $\lambda=0$ and $\epsilon<0$, Eq.~(\ref{flowalpha}) reduces to 
$$
\frac{d\alpha}{dl} =\alpha\left(-\epsilon-\chi+ \frac{\epsilon}{3} \right)=
\alpha\left(-\chi-\frac{2\epsilon}{3} \right)\,$$
leading to $\chi={2(4-d)}/{3}$. 

\begin{figure}
\includegraphics*[width=3cm]{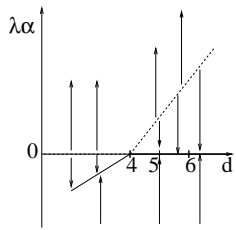}
\caption{The flow of the coupling $\lambda\alpha$ near $d=4$
      dimensions. The arrows indicate the direction of the RG flow of
$\lambda\alpha$. The dashed line marks unstable fixed points, and the
solid line stable fixed points. 
\label{fig3} }
\end{figure}

Let us next discuss the fixed point with $\lambda \neq 0$.
A nonzero $\lambda$ requires $\chi=0$. 
The combination  $\alpha\lambda$  then acts as an effective coupling, and Eq.~(\ref{flowalpha})
has a non-trivial fixed point at
\begin{equation}
\alpha \lambda = \epsilon \frac{K \nu (K+\nu)}{D_m K_4} + {\cal O} (\epsilon^2),
\end{equation}
for $\epsilon>0$ and a  fixed point
\begin{equation}
\lambda\alpha = \epsilon \frac{2K \nu (K+\nu)}{3 D_m K_4} + {\cal O} (\epsilon^2),
\end{equation}
for $\epsilon < 0$. The couplings $a,$ $b$, and $c$ have scaling
dimension zero (to order $\epsilon$) and are thus marginal at this fixed point. 
Determination of their marginal relevance or irrelevance requires 
evaluation of higher order terms in $\epsilon$, which was not attempted in this paper.  
For $\epsilon<0$, the fixed point is {\em stable} as indicated by the flows sketched  in Fig.~\ref{fig3}. 
We expect it to correspond to a rough phase, with the roughness exponent $\chi=0$,
possibly receiving corrections in higher order in $\epsilon$.
For $\epsilon>0$, the fixed point can be interpreted as describing a
roughening transition. The fixed point is unstable (see Figure
\ref{fig3}), with flows to either a flat phase (if the initial
$\lambda$ value is smaller than the fixed point value) or to a rough
phase, which is not accessible by perturbation theory. Compared to a
system that is described by the height variable alone, without a
coupling to critical order parameter fluctuations
\cite{kar86}, the lower critical dimension for the roughening transition
is increased, at criticality of order parameter fluctuations, from 2 to 4.
We thus have found that the coupling to the critical order parameter
fluctuations changes the scaling behavior of the height variable below
6 dimensions, and that it increases the lower critical dimension for
the roughening transition from 2 to 4. Studying the influence of the
surface roughness on the critical order parameter fluctuations near 4
dimensions was not possible to us within perturbation theory.  To
order $\epsilon$, the parameters $a$, $b$, $c$ are marginal, 
and to higher orders in $\epsilon$ the number of
diagrams becomes large. Furthermore, it is not clear whether further
fixed points except those discussed by us are at all accessible by
perturbation theory.

In the case of the conserved Langevin Eqs.~(\ref{langevin3}) and
(\ref{langevin4}) proposed for MBE growth, the mutual effects
between height and order parameter are much weaker. The Gaussian fixed
point is stable above $d=2$ dimensions. Below $d=2$, the coupling
$\beta$ becomes relevant. If $K>0$ and $v$ is small ($v\ll
\Lambda^2/K$), the diffusion of the order parameter on the surface affects the
height profile. Performing an analysis analogous to the one above near
6 dimensions (where $\alpha$ was the relevant coupling), we find a
stable fixed point in $d=2-\epsilon$ with
$\beta^2=D_h\kappa^3\pi\epsilon/2 D_m^2 + {\cal O}(\epsilon^2)$ and
$\chi=2-d$. (Without coupling to the order parameter, there is a
smaller roughness exponent of $\chi=(2-d)/2$.)
For negative $K$, there can be other interesting (non critical)
effects, as mentioned before.

\section{Computer simulations}
\label{simulations}

\begin{figure}
\includegraphics*[width=5.5cm]{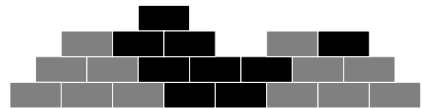}
\caption{The ``brick wall'' model used in the simulations.\label{fig4} 
}
\end{figure}

While the RG gives information about scaling
behavior in high dimensions (4 and higher), computer simulations are
particularly efficient in low dimensions. In this section, we
present results  from simulations in 1+1 dimension. 
Rather than discretizing
Eqs.~(\ref{langevin1}-\ref{langevin2}) and integrating them
numerically, we  perform numerical studies of a
``brick wall'' restricted solid-on-solid model (see
Fig.~\ref{fig4}). 
Since this model shares the same symmetries and
conservation laws as the Langevin equations, it 
should share some of the same universality properties.
However, as noted by Oskoee, Khajehpour, and Sahimi \cite{OKS}
there are also important distinctions between the continuum and discrete model.
In the continuum model at zero temperature, the coarsening force on
a domain of length $\ell$ decays exponentially with $\ell$. 
The corresponding coarsening time then grows logarithmically in $\ell$
($z\to\infty$) \cite{rut94}, as opposed to $z=2$ for Glauber dynamics \cite{gla63}.

Starting from a flat surface, particles
are added such that no overhangs are formed, and with the center of
each particle atop the edge of two particles in the layer below.  We
use two types of particles, $A$ and $B$ (black and grey in the
figure).  The probability for adding a particle to a given surface
site, and the rule for choosing its color, depend on the local
neighborhood.  Since growth is slower on slopes, these growth rules
correspond to $\lambda<0$ \cite{kru90,bar95}.

When $A$ ($B$) particles are more likely to be
added to $A$ dominated regions ($B$), the particles tend to
phase separate and form domains.  In this case, the order parameter
correlation length $\xi$ is of the order of the average domain width.
By changing the growth rules, it is possible to study cases in which
some (or all) of the couplings $a$, $b$, $c$, and $\alpha$ vanish, and
thus to gain a more complete picture of the different ways in which the
height and the order parameter influence each other.

When all the couplings between the order parameter and the height
vanish ($a=b=c=\alpha=0$), the well-known critical exponents $z_h=3/2$
and $\chi=1/2$ of the KPZ equation \cite{kar86,hal95}, and 
$z_m=2$ and 
 $\eta=1$ of the Glauber model \cite{gla63} are recovered. 
This situation is implemented in the following way: A
surface site is chosen at random, and a particle is added if it does
not generate overhangs.  Its color is then chosen depending on the
colors of its two neighbors in the layer below.  If both neighbors
have the same color, the newly added particle takes this color with
probability $1-p$, and the other color with probability $p$ (where $p$
is much smaller than 1).  If the two neighbors have different colors,
the new particle takes either color with probability 1/2.  Neighbors
within the same layer are not considered. As discussed in
Ref.~\cite{ourprl}, these growth rules lead to an order parameter correlation
length $\xi\sim 1/\sqrt{p}$ as $p\to0$.

Here, we want to focus on the more interesting situations where either
$\alpha$ or $a$, $b$, $c$ (or all of them) are nonzero. In the first case
presented below,
the order parameter affects the height variable, but is not influenced
by it. In the second case, the height profile affects the order
parameter dynamics, but not vice versa. One would expect that in the
first case the order parameter imposes its dynamical exponent $z=2$ on
the dynamics of the height profile, and that in the second case the
height profile imposes its exponent $z=3/2$ on the order
parameter. This latter is, however, not the case for $a/\lambda<0$, and we
shall see that $z_m$ is nonuniversal in this case. In the fully coupled
case with $\alpha$, $a$, $b$, and $c$ not equal to zero, we find  $z=2$ or $z=3/2$,
depending on the sign of $(\lambda\alpha)$. 
Some of these results were already reported
in Ref.~\cite{ourprl}. 

\subsection{Growth influenced by independent
phase ordering ($\alpha\neq 0$, $a$, $b$, $c=0$)}

The situation $\alpha>0$ ($\alpha<0$) is implemented by updating sites
on top of particles of different colors less (more) often by a factor
$r<1$ ($r>1$) compared to sites above particles of the same color.  If
the color of the new particle depends only on the neighbors in the
layer below, the order parameter is not affected by the height
variable, and its dynamics is still the same as that of an Ising model,
with $z_m=2$.

\begin{figure}
\includegraphics*[width=6cm]{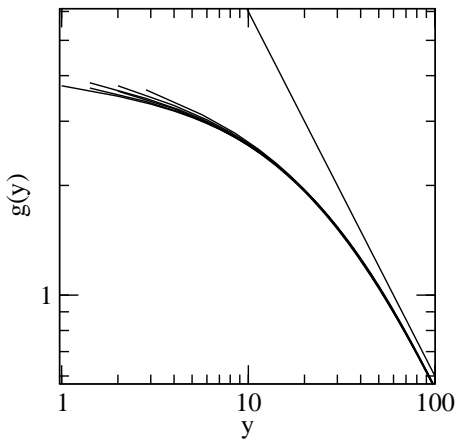}
\caption{The scaling function $g(y)$, obtained for $L=2048$ 
by collapsing data for
$p=320$, 640, 1280, and 2560. For $r$, the values 0.05 and 0.025 were
used. The dashed line is a power law $\propto 1/y$. \label{fig5} 
}
\end{figure}

We first discuss the case $\alpha>0$: Because growth is slower at
domain boundaries than within domains, the domain boundaries sit
preferentially at the local minima of the height profile, with a mound
over each domain. This implies that the surface roughness exponent is
$\chi=1$ on length scales up to $\xi$.  Changes in the height profile
on this scale result from domain wall diffusion, and the dynamical
exponent is therefore $z_h=2$.  On length scales much larger than
$\xi$, the average order parameter is zero, implying that KPZ
exponents of $\chi=1/2$ and $z_h=3/2$ are regained.  The crossover in
the roughness is described by a scaling form
$$ \langle[ h(x,t)-h(x',t)]^2\rangle = |x-x'|^2 g\left(|x-x'|\over \xi\right),$$ 
with a constant $g(y)$ for $y\ll 1$, and $g(y) \sim 1/y$ for $y \gg 1$.
Figure \ref{fig5} shows our simulation results for $g(y)$, obtained by
the data collapse of
$ \langle[ h(x,t)-h(x',t)]^2\rangle/|x-x'|^2 $ versus $|x-x'|\sqrt{p}$,
for different values of $p$.
The curves for the two different values of $r$ differed only slightly, and
were collapsed by multiplying the curves for $r=0.05$ by a factor of
1.06. The scaling collapse is compatible with $g(y) \sim1/y$ for large
$y$, and $g(y) \to$constant, for small $y$. 

For $\alpha<0$, growth occurs with a larger probability at domain
boundaries. Therefore, domain boundaries sit at local maxima. However,
further away from the domain boundaries, their effect is not felt,
and we find $\chi=1/2$ and $z_h=3/2$, just as in the case $\alpha=0$.

\subsection{Phase ordering influenced by independent
growth ($\alpha=0$;  $a$, $b$, $c \neq 0$)}

The situation $\alpha=0$ is implemented by choosing $r=1$, i.e.,
adding a particle at each possible site with the same probability,
irrespective of the color of its neighbors. 
To mimic the influence of surface roughness on the order parameter
(nonzero $a$, $b$, or $c$ in Eqs.(\ref{langevin2})), the color of a
newly added particle is made dependent not only on those of its two
neighbors in the layer below, but also on the colors of its two
nearest neighbors on the same layer, if these sites are already
occupied.  With probability $1-p$, the newly added particle takes the
color of the majority of its 2, 3, or 4 neighbors, and with
probability $p$ it assumes the opposite color.  If there is a tie, the
color is chosen at random with equal probability.  Since the neighbor
on the hillside of a site is more likely to be occupied than the one
on the valley side, with this rule domain walls are driven
downhill,  corresponding to $a>0$ in Eq.~(\ref{langevin2}).
Also, domains on hilltops can expand more easily than those on slopes
 or in valleys, suggesting a value of $b>0$. 

We reported already in Ref.~\cite{ourprl} that the dynamical critical
exponent $z_m$ associated with the order parameter has a nontrivial
dynamic exponent of $z_m \simeq 1.85$, and not the value $z_m=1.5$, 
which may be expected if the walls  follow the surface fluctuations. 
A potential explanation was provided in Ref.~\cite{ourprb} in connection
with the dynamics of a single domain wall
riding on a growing surface. The growth rules for the surface imply
that sequences of brick addition usually proceed from local minima in the
uphill direction, since the addition of a brick generates a potential
growth site (where a brick can be added without generating overhangs)
at the nearest uphill position.  The walls that try to slide downhill
are therefore faced with an upward avalanche of growth mounds of
different sizes that hamper their downhill motion.

\begin{figure}
\includegraphics*[width=6cm]{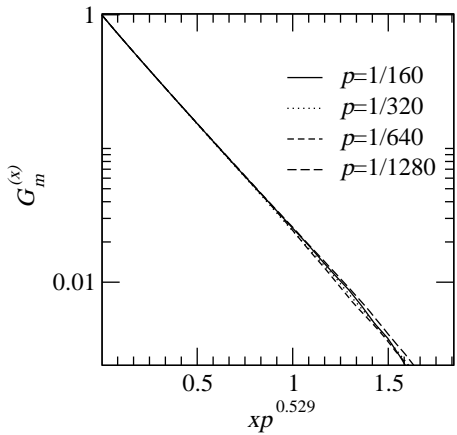}
\caption{Scaling collapse of the correlation functions $G_{m}^{(x)}$
for $q=0.25$ and different values of $p$. 
 \label{fig6} 
}
\end{figure}

\begin{figure}
\includegraphics*[width=6cm]{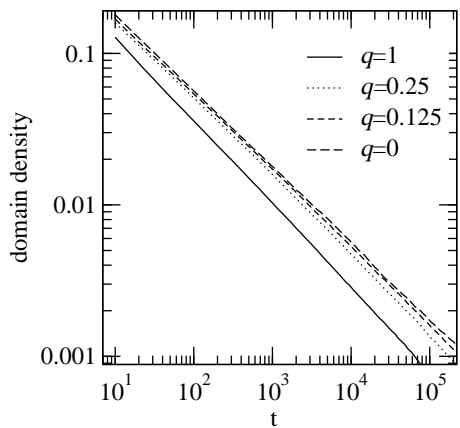}
\caption{Domain density (i.e., number of domains, divided by the system size) as a function of time for $L=16384$,
averaged over 100 samples, for 4 different values of $q$. The curves
are well fitted by power laws of the form $t^{-1/z_m}$ with the
exponents $z_m$=1.8, 1.89, 1.96, 2.0.
 \label{fig7} 
}
\end{figure}

The exponent $z_m$ is not only nontrivial, but also nonuniversal and
depends on the value of $a$. A change in $a$ can be implemented in the
computer simulations by taking into account neighbors within the same
layer as the site being updated with a probability $q$ smaller than 1.
Using the above-mentioned rules and values of $q=1$, 0.25, 0.125, 0,
we find $z_m=1.8$, 1.89, 1.96, 2.0. The difference from the value for
$q=1$ reported in Ref.~\cite{ourprl} stems from the fact that in that
paper we had inadvertently assigned to neighbors within the same layer
a double weight - thus illustrating once more the nonuniversality of
the value of $z_m$. Each of these values was evaluated in two
independent ways, in order to confirm its stability and parameter
dependence. The first was a collapse of the correlation function
$G_m^{(x)}(x)$, as shown in Fig.~\ref{fig6}.
Since $p$ sets the inverse time scale, the scaling variable is
$xp^{1/z_m}$, yielding the value of $z_m$. 
The second method was a domain coarsening simulation,
following a quench from $p=0.5$ to $p=0$. Fig.~\ref{fig7} shows the
number of domains, divided by the system size, as function of time for
different values of $q$.

Equations~(\ref{langevin1}-\ref{langevin2}) allow also for the
possibility of $a<0$. This would imply that domain walls move uphill
and that identical neighbors in the same layer have a positive
interaction energy, in contrast to neighbors in different
layers. Although this is an implausible physical situation, it is
nevertheless of some theoretical interest. For $a=\lambda$ and $c=0$,
the invariance under the transformation $x'=x-\lambda\epsilon t$ and
$h'=h+\epsilon x$ (with a small $\epsilon$) of the KPZ equation holds
also for Eq.~(\ref{langevin2}). 
This invariance corresponds to the Galilean invariance of Burger's equation. 
It implies that the order
parameter dynamics are goverened by the same time scale as the height
variable, i.e.,  $z_m=z_h$. Implementing the case $a<0$ in our
simulations, we indeed find $z_m=1.5$, suggesting that the Galilean
invariant fixed point is the only attractive fixed point in this
domain of parameters.

\subsection{Mutual couplings ($\alpha$, $a$, $b$, $c \neq 0$)}

When all couplings are different from zero, the probability for adding
a particle at a given site and the choice of the particle color depend
on the local neighborhood. The simulation parameter $q$ is positive,
and $r$ is different from 1.  

For $\alpha>0$, we find $z_m=2$ irrespective of the values of
$a$, $b$, and $c$. As particles are added to domain boundaries with a smaller
probability than within domains, domain boundaries sit at local minima
most of the time. Therefore they perform a random walk even when $a
\neq 0$. Over each domain there is a mound, implying that
$\chi=1$ on length scales up to $\xi$.  Changes in the height profile
on this scale result from domain wall diffusion, and the dynamic
exponent is therefore $z_h=2$.  On length scales much larger than
$\xi$, the average order parameter is zero, and KPZ
exponents of $\chi=1/2$ and $z_h=3/2$ are regained.

\begin{figure}
\includegraphics*[width=6cm]{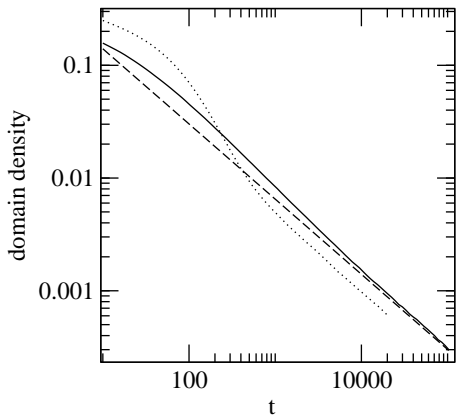}
\caption{Domain density (i.e., number of domains, divided by the
system size) as a function of time for $a\lambda<0$ and $L=16384$,
averaged over 100 samples, for $q=1$ and $r=5$ (solid line). The
dashed line is a power law with an exponent $-2/3$, corresponding to
$z_m=1.5$. The dotted curve shows the corresponding  simulation result with positive $a\lambda$.
\label{fig8} }
\end{figure}

For $\alpha<0$, particles are added rapidly on domain boundaries, and
domain walls can therefore not be trapped in local height minima. For the case $a\lambda>0$, where domain walls tend to move uphill, we therefore expect that the situation 
$\alpha<0$ is similar to the case $\alpha=0$, for which we found $z_m=z_h=3/2$. 
This means that the above-mentioned Galilean-invariant fixed point remains applicable to $\alpha < 0$. 
The simulation results shown in Fig. \ref{fig8} confirm this expectation. 
For the case $a\lambda<0$ and $\alpha=0$, we have argued above that 
the downhill motion of domain walls is hampered by an upward avalanche of 
growth mounds, which cause the walls to be temporarily stuck in local minima, 
leading to a nonuniversal exponent $z_m$. 
Now, for $\alpha<0$ and $a\lambda<0$, we find in our simulations that the 
dynamical critical exponent $z_m$ is identical to
$z_h=1.5$, implying that the downhill motion of the domain walls 
is not hampered any more but that the domain walls can follow the height fluctuations. 
Fig.~\ref{fig8} shows the results of a domain coarsening
simulation for the parameters $q=1$ and $r=5$. For comparison,
simulation results for positive $a\lambda$ (and otherwise the same parameter values) are also shown. One can see that the exponent $z_m$ is indeed the same in both situations.
To summarize, we find that for $\alpha<0$ the height profile imposes
its critical behavior on the order parameter fluctuations, while in
the opposite case, $\alpha>0$, the domain wall diffusion imposes a
dynamical exponent $z=2$ on the system. 

\subsection{Comparison to analytical results}

It is interesting to compare the results of the computer simulations 
with the admittedly limited results of the RG analysis presented in
the previous section. 

Let us start from our last result that for $\alpha\lambda>0$ the
height profile imposes its critical behavior on the order parameter,
while the reverse is true for $\alpha\lambda<0$. The RG up to order
$\epsilon=4-d$ showed that 
for 
$\alpha\lambda<0$ the fixed point is
accessible perturbatively, with $z=2$ to order $\epsilon$. It is
striking that this result holds also in $1+1$ dimension. 
For $\alpha\lambda>0$, the RG flow runs away to infinity, 
suggesting the existence of a strong coupling fixed point. 
The critical behavior corresponding to this fixed point was  found in 
our simulations in 1+1 dimension to be the same as that of the KPZ equation, 
with $z=3/2$. If this result is not
particluar to 1+1 dimension, it suggests that a positive
$\alpha\lambda>0$ might be irrelevant at the KPZ strong coupling fixed
point. However, this conclusion can not be tested via by
perturbative  RG analysis.

Up to order $\epsilon$ of the RG analysis, the parameters $a$, $b$, $c$
are marginal.  It appears from our
simulations that these three coupling indeed do not modify the
critical behavior as long as $\alpha
\neq 0$, suggesting that these parameters are marginally irrelevant.

For $\alpha=0$, the critical behavior of the height profile is given
by the KPZ equation, and it has a stable fixed point at $\lambda=0$ in
which the dimension of $h$ is $(2-d)/2<0$. Thus at the weak coupling
fixed point, $a$, $b$, and $c$ are irrelevant in four dimensions and
the Ising fixed point is not modified by coupling to the height
parameter. However, the RG analysis cannot predict the critical
behavior of the coupled system in the rough phase, which is
characterized by a strong coupling fixed point. Nevertheless, we could
argue that there exists a Galilean-invariant fixed point when $a\lambda>0$,
where the
height profile imposes its dynamical critical exponent on the order
parameter. The result $z_m=2$ in 1+1 dimensions was confirmed by our
computer simulations, suggesting a larger domain of parameter space
where such scalings apply.
 
 The only case for which we have no analytical prediction is when $\alpha=0$, and
$\lambda a <0$, where the computer simulations reveal nonuniversal behavior.

\section{Relation to passive scalar advection and drifting polymers}
\label{other}

There is a close connection between the model studied in this paper
and other coupled nonequilibrium systems. One such system is obtained
when we regard domain walls as ``particles'' that 
ride on the growing surface. 
With the substitution $\nabla m = \rho$, we obtain for 
 $u=b=c=0$ the following  equation for $\rho$:
\begin{equation}
\frac{\partial \rho}{\partial t} = {\kappa} \nabla^2\rho+{a}\nabla
(\rho \nabla h)+\zeta_{\rho}({\bf x},t), \label{rho}
\end{equation}
with a conserved noise $\zeta_{\rho}$. Without the noise term this equation is the Fokker Planck equation corresponding to 
the Langevin equation 
\begin{equation}
\frac{d {\bf x}}{dt} = a\nabla h + \zeta_x(t)\,.
\label{advection}
\end{equation}
If we assume that the particles do not interact with each other, 
the Fokker Planck equation, combined with a noise term, describes the time evolution of the particle density. 

If we require additionally that there is no effect of the
particles on the growing interface, the coupling $\alpha$ vanishes,
too.  The substition $\nabla h = \vec v$ then turns the KPZ equation
into a randomly stirred Burger's equation, and Eq.~(\ref{advection})
becomes the equation of motion of a particle advected by the flow.
This model was studied in detail in Ref.~\cite{ourprb}. Just as for the
model described in this paper, the scaling behavior of the advected
particles is fundamentally different in the two cases $a/\lambda>0$
and $a/\lambda < 0$. In the first case, the system has a Galilean
invariant fixed point, and particle diffusion is characterized by a
dynamical critical exponent $z_\rho=3/2$ in one dimension, while this
exponent is larger than $3/2$ and nonuniversal in the other case.

Finally, Eqs.~(\ref{langevin1}-\ref{langevin2}) with $a=b$ and
$c=r=u=0$, but with $\alpha \neq 0$, can be mapped on the equations
used to describe the dynamics of a streched string moving in a random
medium \cite{ert92}. 
If the string is streched in the $x$-direction, and if it
moves in the $h$-direction, the configuration of a string embedded in
3 dimensions can be characterized by the coordinates $h(x)$ and
$h_\perp(x)$. Assuming that the evolution of the line is dissipative
and local, the equations of motion then are our equations 
Eqs.~(\ref{langevin1}-\ref{langevin2}) with the replacement $m=\partial_x h_\perp$ and with $\zeta_m$ replaced with a conserved noise $\partial_x \zeta_m$.
Apart from the Galilean-invariant fixed point, this set of equations has a
fixed point where a fluctuation-dissipation relation is satisfied
(i.e. where a stationary solution of the Fokker-Planck-equation can be
written down). 
This fixed point corresponds in our notation to the
situation $\alpha=Ka/\nu$.
(For a general discussion of equations the stationary solutions of which can 
be calculated, see Ref.~\cite{rava}).

\section{Conclusions}
\label{conclusions}
In summary, the interplay between surface roughening and phase
separation leads to a variety of novel critical scaling behaviors.  At
one extreme, the height profile adapts to the dynamics of critical
domain ordering. At the other extreme, the dynamics of the domain
walls follow the height fluctuations. For a third range of paramter
values, the dynamics of domain wall motion is influenced by the
roughness, but exhibits nontrivial and nonuniversal scaling behaviors.

Several generalizations of the model presented in this paper are
possible. For example, as discussed in Ref.~\cite{kar99}, one can consider
the situation where the symmetry breaking involves a continuous,
rather than an Ising-like order parameter. Such a situation applies to
the deposition of spins that can realign on the surface but are frozen
in the bulk, or to orientational symmetry breaking in the plane during
crystal growth.
Another generalization would be the inclusion of elastic forces, which
are often present during the growth of composite films (see Ref.~\cite{rava}). 
Furthermore,
one could consider phase transitions where the different types of
molecules order on sublattices instead of phase separating. 

Finally,
there exist growth situations where one type of particles is
magnetic. In addition to a ordering or phase separation transition
which occurs at the surfaces, there is in this case also a magnetic
phase transition, which occurs in the bulk. This combination of
two-dimensional and three-dimensional phase transitions is
particularly challenging for a theoretical analysis, and it leads to
interesting experimental results \cite{hel95}.

\begin{acknowledgments}
This work was supported by the Deutsche 
Forschungsgemeinschaft (DFG) under Contract No Dr300-2/1 (BD),
and by the National Science Foundation grant DMR-01-18213 (MK).
 \end{acknowledgments}

\end{document}